\documentclass{aa}
\usepackage{psfig}
\def\la{\;
\raise0.3ex\hbox{$<$\kern-0.75em\raise-1.1ex\hbox{$\sim$}}\; }
\def\ga{\;
\raise0.3ex\hbox{$>$\kern-0.75em\raise-1.1ex\hbox{$\sim$}}\; }
\newcommand{\zabs}{$z_{\rm abs}\,$}
\newcommand{\hh}{H$_2\,\,$}
\newcommand{\te}{$T_{\rm eff}\,$}
\newcommand{\tr}{$T_{\rm CMBR}\,\,$}
\begin{document}
\title{The cosmic microwave background radiation temperature
at \zabs = 3.025 toward QSO 0347--3819\thanks{Based on public
data released from UVES Commissioning at the VLT Kueyen telescope,
ESO, Paranal, Chile}
}
\author{ P. Molaro\inst{1}
\and
S. A. Levshakov\inst{2}\thanks{On leave from the
Ioffe Physico-Technical Institute, Russian Academy of Sciences,
St.~Petersburg}
\and 
M. Dessauges-Zavadsky\inst{3,4}
\and 
S. D'Odorico\inst{3}
}
\offprints{P. Molaro}
\institute{
Osservatorio Astronomico di Trieste,
Via G.B. Tiepolo 11, I-34131 Trieste, Italy\\
%\email{molaro@oat.ts.astro.it}
\and
Division of Theoretical Astrophysics,
National Astronomical Observatory, Mitaka, Tokyo 181-8588, Japan\\
%\email{lev@astro.ioffe.rssi.ru}
\and
European Southern Observatory,
Karl-Schwarzschild-Str. 2,
D-85748 Garching bei M\"unchen, Germany\\
%\email{sdodoric@eso.org}
\and
Observatoire de Gen\`eve, CH--1290 Sauverny,
Switzerland\\
%\email{ mdessaug@eso.org}
}
\date{Received 00  / Accepted 00 }

\abstract{
From the analysis of the C$^+$ fine-structure  population ratio
in the damped Ly$\alpha$ system at \zabs = 3.025
toward  the quasar
Q0347--3819 we derive an upper bound of $14.6\pm0.2$ K 
on the cosmic microwave background
temperature (\tr) regardless the presence of other
 different excitation 
mechanisms.
The analysis of the ground state 
rotational level populations of \hh detected in the system
reveals a Galactic-type  UV radiation field ruling out UV 
pumping as an important excitation mechanism for C$^+$. 
The low dust content estimated from the Cr/Zn ratio indicates
that the IR dust emission can also be neglected.
When the collisional excitation is considered,
we measure a  temperature for the cosmic background radiation of
\tr =$12.1^{+1.7}_{-3.2}$ K.
The results are in  agreement with 
the \tr = $10.968$ $\pm0.004$ K predicted by  
the hot Big Bang cosmology  at \zabs = 3.025. 
}
\titlerunning{\tr at \zabs = 3.025}
\authorrunning{Molaro, Levshakov, Dessauges-Zavadsky \& D'Odorico}
\maketitle

\section{Introduction}

\addtocounter{footnote}{4}

In the standard Big Bang model (SBB)
the temperature of the relic radiation  from
the hot phase of the Universe is predicted to
increase linearly with redshift:
\tr$(z)$ = \tr$(0)\,(1+z)$ 
(e.g., Peebles 1993).
At the present epoch direct measurements show that
\tr(0) =
$2.725\pm0.001$~K ($1\sigma$ c.l.), and that the relic radiation follows
a Planck spectrum with very high precision (Mather et al. 1999). 

As pointed out by Bachall \& Wolf (1968) the  CMBR temperatures at
earlier epochs can be measured from the
analysis of quasar absorption line spectra which show 
atomic and/or ionic fine-structure levels
excited by the photo-absorption of the CMBR.
Among the species with fine structure levels 
the \ion{C}{i} and \ion{C}{ii}
show an  energy separation, 
from 23.6 K up to 91.3 K,  which make them 
sensitive to the CMBR, in particular as the 
redshift increases.
However,  \ion{C}{i} is generally fully 
ionized and rarely detected, while the 
\ion{C}{ii} ground-state transitions are strongly saturated, 
thus  making column densities 
rather uncertain. 
In addition, non cosmological sources
(such as particle collisions, pumping by UV radiation,
IR dust emission and by other sources)
may compete with the CMBR to populate the excited 
fine-structure levels. 
Only independent knowledge of ambient radiation field and of  particle
densities
allows to disentangle the  contribution of the background radiation
from that of other mechanisms.
For these reasons  previous measurements  
place upper limits to \tr rather than real measurements, albeit 
quite stringent ones 
(Meyer et al. 1986; Songaila et al. 1994; Lu et al. 1996; 
Ge, Bechtold \& Black 1997;
Roth \& Bauer 1999; Ge, Betchold \& Kulkarni 2001).

Recently, Srianand, Petitjean \& Ledoux (2000)  from the
\hh  analysis in the DLA at  
\zabs = 2.3371 toward the quasar Q1232+0815 were  able to 
infer the UV radiation field in the absorber. 
Then by means of \ion{C}{i},  
\ion{C}{i}$^\ast$, \ion{C}{i}$^{\ast \ast}$,
\ion{C}{ii} and \ion{C}{ii}$^\ast$ they 
obtained  a \tr  = $10\pm4$ K, while
SBB predicts \tr = $9.09$ K. However,
 the H$_2$ abundance measurement at \zabs = 2.3371 by
Srianand et al. (2000) is in contradiction with their recent
estimation of the deuterated molecular hydrogen  abundance
(Varshalovich et al. 2001). The ratio
HD/H$_2$ $\simeq (0.8\div3.0)\times10^{-3}$, whereas it is $10^{-7}\div10^{-6}$
 in the ISM diffuse clouds 
(e.g. Wright \& Morton 1979). Until this discrepancy is 
clarified  the
value of Srianand et al. (2000)  should be taken as
an upper limit of $T_{\rm CMBR} < 14$~K  at \zabs = 2.3371.  
So far all  measurements have been found to be consistent 
with the SBB model prediction.

In this letter, we present  a new measurement  of \tr
at higher redshift, \zabs = 3.025, from the VLT/UVES
spectra of Q0347--3819.

\section{Data analysis and results}

The spectroscopic observations of Q0347--3819 obtained during  
UVES  commissioning at the VLT 8.2 m telescope 
are described in detail
by D'Odorico, Dessauges-Zavadsky \& Molaro (2001) and by
Levshakov et al. 2002 (LDDM, hereinafter).

\begin{figure}
\vspace{0.0cm}
\hspace{0.0cm}\psfig{figure=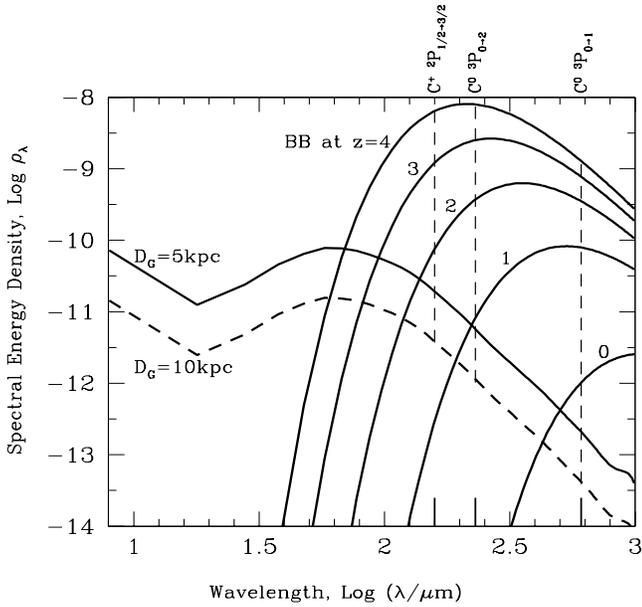,height=10.0cm,width=8.0cm}
\vspace{-1.5cm}
\caption[]{The Galactic interstellar radiation fields at
galactocentric distances $D_{\rm G} = 5$ and 10 kpc from
 Mathis et al. (1983) are compared with  the black body spectra
calculated at different redshifts using the  relation
\tr$(z)$ = \tr$(0)\,(1+z)$. 
The positions of the excited levels
of C$^0$ and C$^+$, suitable to constrain the \tr values
at different redshifts,  are indicated by vertical  dashed lines
}
\end{figure}

In  LDDM relevant  physical properties for
the damped Ly$\alpha$ system (DLA)
at \zabs = 3.025 are obtained by analyzing
numerous \hh and metal absorption lines associated to the DLA.
The $z_{\rm abs} = 3.025$ system  exhibits 
a multicomponent velocity structure
spanning  over $\sim 80$ km~s$^{-1}$.
The main component at \zabs = 3.024855 has  
a hydrogen column density of 
$N$(\ion{H}{i}) = $2.52(\pm0.04)\times10^{20}$ cm$^{-2}$ and shows the
presence of molecular hydrogen with 
a fractional abundance of 
$f({\rm H}_2) = 3.25(\pm 0.17)\times10^{-6}$.
Several neutral and ionized species associated with this  
cloud have been analyzed. 
In particular the \ion{C}{ii} 1036.3367 and  
\ion{C}{ii}$^\ast$ 1037.0182 lines have been identified.

A line absorption model for the \zabs = 3.025 system  
which is able to reproduce the
line profiles  for  the whole set of atomic and ionic species 
has been elaborated in LDDM. 
This model allows to get a reliable column density
of the saturated lines. LDDM obtained a 
column density of  
$N$(\ion{C}{ii}) = $5.05(\pm0.28)\times10^{15}$ cm$^{-2}$
for the \ion{C}{ii} 1036.3  main component.
This C column density is consistent with what can be inferred 
from the other elements measured in the 
system by means of unsaturated lines.
For instance, if C goes in lockstep  with the undepleted \ion{Zn}{ii} 
we would obtain  $N$(\ion{C}{ii})= 3.1$\times 10^{15}$ cm$^{-2}$, assuming solar photospheric
 values from Grevesse and Sauval (1998),
while we  would obtain   $N$(\ion{C}{ii})= $5.8\times10^{15}$ cm$^{-2}$ 
if C follows \ion{Ar}{i}, with the Ar solar value quoted in 
Sofia \& Jenkins (1998). Neutral carbon is not detected  and  
N(\ion{C}{i})/N(\ion{C}{ii})  $< 7.9\times10^{-5}$.

The column density
for the  $N$(\ion{C}{ii}$^\ast$)  1037.0182 main component is 
$2.26(\pm 0.12)\times10^{13}$ cm$^{-2}$.
Prochaska \& Wolfe (1999)  reported the detection of the
\ion{C}{ii} 1334.5323 and  \ion{C}{ii}$^\ast$ 1335.7077 lines
in the same system. For the latter line,
which is unsaturated, they provide a column density   
of $N$(\ion{C}{ii}$^\ast$) = $3.00(\pm 0.23)\times10^{13}$ cm$^{-2}$, 
which refers to the total system. When we correct 
this value according to the relative ratios between the  
components [$1:0.195:0.044:1.952$ (LDDM)], 
we obtain  for the main one
$N$(\ion{C}{ii}$^\ast$) = $1.83(\pm0.14)\times10^{13}$ cm$^{-2}$.  
The  \ion{C}{ii}$^\ast$ 1335.7077 line is likely  blended with the 
\ion{C}{ii}$^\ast$ 1335.6627 which produces
the blue asymmetry present in the Keck spectrum at --10 km s$^{-1}$ 
(cf. Fig. 5 in Prochaska \& Wolfe).
The relative strengths of the two blended transitions 
is $\it f_{1335.7}/f_{1335.6}$ = 8.7. If  we correct
$N$(\ion{C}{ii}$^\ast$) by the corresponding factor, 
for optically thin lines we obtain 
$N$(\ion{C}{ii}$^\ast$) = $1.64(\pm0.11)\times 10^{13}$ cm$^{-2}$ 
(main component).
The weighted mean between the VLT and Keck
quantities is $N$(\ion{C}{ii}$^\ast$) = $1.92(\pm0.08)\times10^{13}$
cm$^{-2}$.
Combining this value with the ground level 
column density obtained from the VLT
we derive a ratio 
$N$(\ion{C}{ii}$^\ast$)/$N$(\ion{C}{ii}) = $3.8(\pm 0.3 )\times10^{-3}$.

The ground state of the C$^+$ ion consists of two levels
$2s^2 2p$~$^2$P$^0_{1/2,3/2}$  with an energy separation of 
$\Delta E = 63.42$ cm$^{-1}$
which corresponds to $\lambda = 157.7$ $\mu$m.
The excited level can be populated by several mechanisms 
such as collisions,
fluorescence or IR photon absorption, which include also  the CMBR.
In the following we use an effective temperature $T_{\rm eff}$
 to characterize at 
$\lambda =157.7$ $\mu$m the proper 
spectral energy density of the local IR field 
approximated by a Planck spectrum with $T = T_{\rm eff}$.

In equilibrium,
the population ratio of the upper level $n_2$ to the lower level
$n_1$, in
ions with a doublet fine structure in the ground state,
is given by:
\begin{equation}
\frac{n_2}{n_1} = \frac{Q_{1,2} + w_{1,2}}{
Q_{2,1} + A_{2,1} + w_{2,1}}\; ,
\label{eq:E1}
\end{equation}
where $w_{1,2}$, $w_{2,1}$ and $Q_{1,2}$, $Q_{2,1}$ are
the photo-absorption, radiative decay, collisional
excitation and de-excitation rates, respectively.
$A_{2,1}$ is the radiative transition probability which is 
$A_{2,1} = 2.291 \times 10^{-6}$ s$^{-1}$ for C$^+$.

If only the background radiation contributes to the population 
of the excited fine-structure states, eq.
(1) gives: 
\begin{equation}
T_{\rm CMBR}(z) = \frac{91.325} 
{\ln\, \left[g^\ast\,N(CII)/g\,N(CII^*)\right]}\; ,
\label{eq:E2}
\end{equation}
where $g^*$ and  $g$ are the statistical weights of the
corresponding levels.
Thus the estimated $N$(\ion{C}{ii}$^\ast$)/$N$(\ion{C}{ii}) ratio
would yield \te = $14.6\pm0.2$ K, 
which can be considered as a firm upper limit to the \tr.  
The  value  expected from the standard model at \zabs $ = 3.025$ is of
\tr = $10.968\pm0.004$ K. 
Thus the observed  
$N$(\ion{C}{ii}$^\ast$)/$N$(\ion{C}{ii}) ratio  provides a 
stringent
limit to the \tr at \zabs $ = 3.025$ with only little room  left for
extra contributions. 

In the following we show  how the detection of H$_{2}$ in the
same component  where  \ion{C}{ii}$^\ast$ and \ion{C}{ii} are observed
can provide additional information on the presence of other excitation
processes.  In this discussion  we assume 
that  the molecular and ionic species trace 
the same material as it is suggested by the 
similar broadening shown by the line profiles
and by the absence of any evidence for an associated  
dense \ion{H}{ii} region gas on the line of sight
as argued in LDDM from the non detection of 
the \ion{N}{ii}$^\ast$ 1084.580 and 1084.562 lines. 
 
The \hh populations over
the J = 0 to J = 5 rotational levels of the ground 
electronic-vibrational state provide
an excitation temperature of $T_{\rm ex}$ = 825 $\pm$ 110 K and the 
kinetic temperature is also estimated to be  
$T_{\rm kin} \la 430 $ K (LDDM).
The population ratios of the higher J levels 
$N$(5)/$N$(3) and $N$(4)/$N$(2) are 
sensitive to the UV pumping. The measured rate of photo-absorption 
$\beta _0$ $\approx 2 \times 10^{-9}$ s$^{-1}$ is 
very close to the average 
interstellar radiation field in the Galaxy.
With this constraint on the UV flux  
the fluorescent 
excitation process has a  rate  $\simeq
9.3\times10^{-11}$ s$^{-1}$, which is rather low and can be neglected
according to Silva \& Viegas (2001).

The rates of the radiative processes $w_{1,2}$
and $w_{2,1}$ may be caused by 
the cosmic microwave background radiation at \zabs = 3.025,
but also by local sources of infrared radiation like
diffuse emission from dust heated by OB stars
to temperatures
$T  \simeq 10 \div 20$~K as observed in the Milky Way
(Mathis, Mezger \& Panagia 1983).
The possible contribution from the heated dust is 
illustrated in Fig.~1 where
the MW interstellar radiation fields at galactocentric distances
$D_{\rm G} = 5$ and 10 kpc are shown along with the black body
spectra calculated at different redshifts using the linear relation
\tr$(z)$ = \tr$(0)\,(1+z)$. 
This is representative of our system since, 
as we have
discussed above, the intensity of the UV field in the \zabs =3.025 
cloud is found 
to be of the same order of magnitude as in the MW.
The positions of the excited levels
of C$^0$ and C$^+$, which are suitable to restrict \tr,
are also indicated by vertical lines. 
Fig.~1 shows  that the diffuse FIR reemission
of stellar radiation by dust grains, if the dust emissivity at
$\lambda = 157.7$ $\mu$m is equal to the highest value measured at
$D_{\rm G} = 5$ kpc in the MW, always remains lower than the 
expected CMBR. 
The  corresponding photo-absorption
rate is $w^{\rm dust}_{1,2} \simeq 1.8\times10^{-11}$ s$^{-1}$,
but the expected rate induced by the relic radiation is
$w^{\rm CMBR}_{1,2} \simeq 1.1\times10^{-9}$ s$^{-1}$.
Moreover, LDDM estimated  from the [Cr/Zn] abundance ratio that 
the dust content in the \zabs = 3.025 absorbing region
is about 30 times lower as compared with the MW mean value, 
so that we may exclude significant contribution from dust emission.

We now consider the
information on the particle density, since
the upper level of C$^+$ may be
populated by collisions with several particles such as 
electrons, e$^-$, hydrogen atoms,
H$^0$, protons, H$^+$, and molecules, \hh. 
The J=2 level of \hh has a rather long 
radiative lifetime and is the more 
sensitive to the collisional de-excitation. 
The critical density above which collisional de-excitation
becomes important is $ n_{\rm H}^{\rm cr}$=3.6 cm$^{-3}$ (LDDM) and, 
therefore, $n_{\rm H} \ge 4$ cm$^{-3}$ is required 
to maintain the observed $N$(2)/$N$(0) ratio at 
T$_{\rm kin}$ $\simeq$ 400 K.
Arguments based on the production rate of \hh
imply that  the volumetric gas density, $n_{\rm H}$, ranges 
between 4  and 14 cm$^{-3}$.

\begin{figure}
\vspace{0.0cm}
\hspace{0.0cm}\psfig{figure=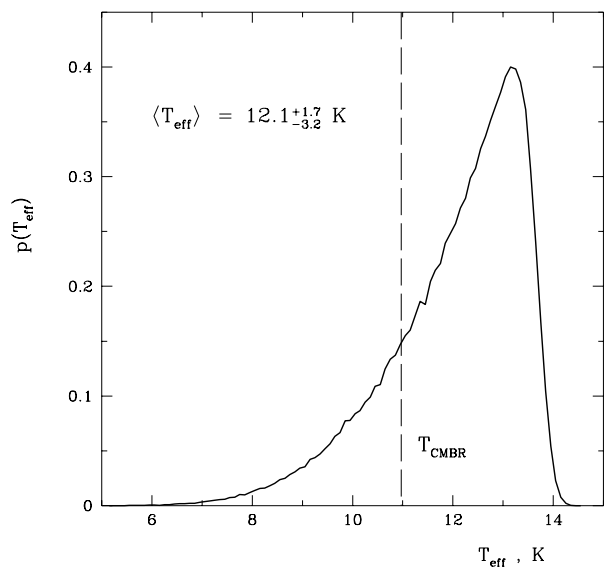,height=6.0cm,width=8.0cm}
\vspace{1.7cm}
\caption[]{
Monte Carlo simulations of the
probability density function of \te\,
for the value of
$N$(\ion{C}{ii}$^\ast$)/$N$(\ion{C}{ii}) = $3.8\times10^{-3}$.
The mean  value is \te\, =
$12.1^{+1.7}_{-3.2}$ K [the $\frac{1}{2}(1-p)$ and $\frac{1}{2}(1+p)$
quantiles were used to estimate the uncertainty interval at 
$p = 0.95$]. 
The \tr from the standard Big Bang cosmological 
model is marked with a vertical dashed line
}
\end{figure}

The H$^0$--C$^+$ collisional rate is of $Q^{{\rm H}^0}_{1,2} 
\simeq 1.45\times10^{-9}\,
n_{\rm H}$\, s$^{-1}$ in the range $10^2$~K $< T_{\rm kin} < 10^3$~K
(Launay \& Roueff 1977).
Collisions with electrons have 
the highest rates but the electron density
is rather low.   
Electrons in \ion{H}{i} regions
come mainly from carbon 
photo-ionization, so that $n_{\rm e} \simeq
{\rm (C/H)}\,n_{\rm H}$,
which is $\simeq 3\times10^{-5}\,n_{\rm H}$ for
the \zabs = 3.025 system. The rate is 
$Q^{{\rm e}^-}_{1,2} \simeq 2\times10^{-7}\,n_{\rm e}$\,
s$^{-1}$ 
and therefore the collisional rate becomes
$Q^{{\rm e}^-}_{1,2} \simeq 6\times10^{-12}\,n_{\rm H}$\,
s$^{-1}$, which is much lower than that of the hydrogen collisions 
for the same temperature interval (Silva \& Viegas 2001).
\hh molecules do not contribute to  
collisions considering the low fractional
abundance measured in the system.
The corresponding de-excitation rate, calculated from  the principle
of detailed balance, is
$Q^{{\rm H}^0}_{2,1} \simeq 9.11\times10^{-10}\,
n_{\rm H}$\, s$^{-1}$, for $T_{\rm kin} \simeq 400$~K. 
 
We calculated the probability density function of
\te\, using statistical Monte Carlo simulations
which suggest that the errors
are normally distributed around the mean value
of $N$(\ion{C}{ii}$^\ast$)/$N$(\ion{C}{ii}) with the
dispersion equal to the probable error of this ratio, while
$n_{\rm H}$ is evenly distributed between
4 cm$^{-3}$ and 14 cm$^{-3}$.
The result is presented in Fig.~2.
The most probable value of
\te\, obtained in this analysis is
\te\, = $12.1^{+1.7}_{-3.2}$~K. The lower and upper errors of \te\, 
correspond to
the $\frac{1}{2}(1-p)$ and $\frac{1}{2}(1+p)$
quantiles, respectively (the central 100p\% confidence
interval was used with $p = 0.95$). 

Since we have considered collisions and excluded 
fluorescence and  dust emission as significant 
processes in the population of the excited levels, 
\te is actually  \tr\ for this particular DLA.
Thus our measurement of
$N$(\ion{C}{ii}$^\ast$)/$N$(\ion{C}{ii})
= $3.8(\pm0.3)\times10^{-3}$ leads to the 
most probable value of 
$T_{\rm eff} = 12.1$~K
which is only  1.1 ~K  higher with respect to the
predicted \tr and fully consistent within errors.

In Fig. 3 all the previous estimations of \tr are shown.
Our result, together with upper limits presented in Fig.~3
support the linear evolution of the CMBR
within the framework of the SBB model.

Alternative non-adiabatic cosmological models in which photon creation
takes place as the Universe expands predict a 
different temperature-redshift relation 
of the type \tr$(z)$ = \tr$(0)\,(1+z)^{(1- \beta)}$ (Lima, Silva \& Viegas 2000).
At high redshift the deviation becomes more pronounced and our measurement set
a limit to $\beta \le$ 0.22 (2 $\sigma$). 
%However a more stringent
 % limit has been derived by 
%Birkel \& Sarkar (1997) by means of primordial nucleosynthesis constraints.

\begin{figure}
\vspace{0.0cm}
\hspace{0.0cm}\psfig{figure=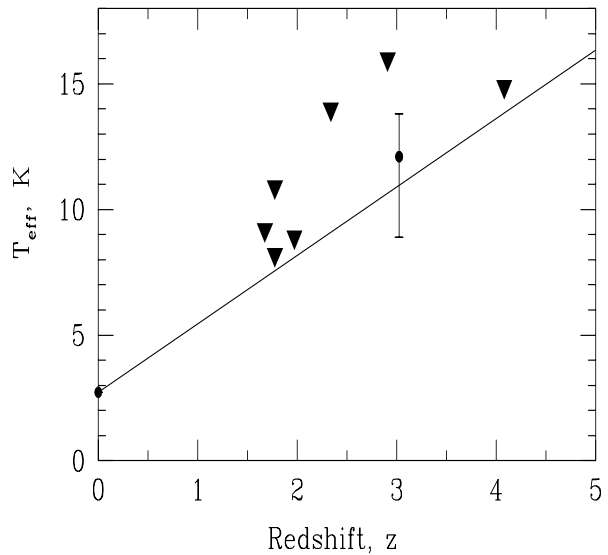,height=10.0cm,width=8.0cm}
\vspace{-2.0cm}
\caption[]{
Measurements of \te at different redshifts. Upper limits from
previous works 
(Songaila et al. 1994; Lu et al. 1996; Ge, Bechtold \& Black 1997;
Roth \& Bauer 1999 and Srianand et al. 2000) 
are marked by triangles. The dot with error
bars shows the estimation of \te\, at \zabs = 3.025 
toward Q0347--3819 (this letter). The straight line shows the
prediction from the hot Big Bang cosmological model,
\tr$(z)$ = \tr$(0)\,(1+z)$
}
\end{figure}

\section{Conclusions}

The  analysis of the \hh lines in the damped Ly$\alpha$ absorber
at \zabs = 3.025  toward  QSO 0347--3819
allows us to estimate the local excitation
mechanisms which populate the fine-structure levels  together
with the \tr. 
From the N(\ion{C}{ii}$^\ast$)/N(\ion{C}{ii}) ratio
we measure the 
temperature of the local background radiation
of \tr\, = $12.1^{+1.7}_{-3.2}$~K which is consistent with
the temperature of the cosmic background microwave radiation
of 10.968 ~K predicted by the 
standard Big Bang cosmology at the redshift
of the absorber.

\begin{acknowledgements}
We thank our anonymous referee for valuable comments and suggestions.
The work of S.A.L. is supported in part 
by the RFBR grant No.~00-02-16007.
\end{acknowledgements}

\end{document}